\documentclass[a4paper, journal]{IEEEtran}
\usepackage{cite}
\usepackage{latexsym}
\usepackage{textgreek}
\usepackage{etoolbox}
\makeatletter
\patchcmd{\@makecaption}
  {\scshape}
  {}
  {}
  {}
\makeatletter
\patchcmd{\@makecaption}
  {\\}
  {.\ }
  {}
  {}
\makeatother
\def\tablename{Table}

\usepackage{tcolorbox}
\usepackage{color}
\usepackage{fontawesome}
\usepackage{amsthm}
\usepackage{amsxtra}
\usepackage[mathscr]{euscript}
\usepackage{graphicx}
\graphicspath{{./fig/}}
\usepackage{multirow}
\usepackage{pifont}

\usepackage{rotating, graphicx}
\usepackage{array,multirow,graphicx}
\usepackage{amsmath}
\usepackage{tikz}
\usepackage{amssymb}
\usepackage{lipsum}
\usepackage{lipsum, multicol}
\usepackage{adjustbox}
\usepackage{algorithm}
\usepackage{algpseudocode}
\usepackage{upgreek}
\usepackage{mathtools}
\usepackage{float}
\usepackage{xcolor}
\usepackage[caption = false]{subfig}
\usepackage{acronym}
\usepackage{orcidlink}
\usepackage{cuted} 
\usepackage{fancyhdr}
\hypersetup{hidelinks}
\usepackage{colortbl,booktabs}
\usepackage{svg}
\usepackage{supertabular}
\usepackage{comment}
\usepackage{alltt}
\usepackage{enumitem}
\usepackage{mathtools}
\usepackage{float}
\usepackage[caption = false]{subfig}
\usepackage{threeparttable}
\usepackage[utf8]{inputenc}
\usepackage[english]{babel}

\makeatother
\usepackage{eso-pic}
\newcommand\AtPageUpperMyright[1]{\AtPageUpperLeft{
 \put(\LenToUnit{0.05\paperwidth},\LenToUnit{-1cm}){
     \parbox{1\textwidth}{\raggedleft\fontsize{9}{11}\selectfont #1}}
 }}
\newcommand{\conf}[1]{
\AddToShipoutPictureBG*{
\AtPageUpperMyright{#1}
}
}

\begin{document}

\title{Network Slicing in 5G Mobile Communication: Architecture, Profit Modeling, and Challenges
\vspace{-0.40mm}}

\conf{This article has been accepted for publication in the International Symposium on Wireless Communication Systems 2017, Bologna, Italy.} 

\author{Mohammad Asif Habibi\orcidlink{0000-0001-9874-0047}, 
Bin Han\orcidlink{0000-0003-2086-2487}~\IEEEmembership{(Senior Member,~IEEE)}, 
and Hans D. Schotten\orcidlink{0000-0001-5005-3635}~\IEEEmembership{(Member,~IEEE)}
\vspace{-11.8mm}
\thanks{Mohammad Asif Habibi\orcidlink{0000-0001-9874-0047}, Bin Han\orcidlink{0000-0003-2086-2487}, and Hans D. Schotten\orcidlink{0000-0001-5005-3635} are with the Division of Wireless Communications and Radio Navigation (WiCoN), Department of Electrical and Computer Engineering (EIT), University of Kaiserslautern (RPTU), $\mathrm{67663}$ Kaiserslautern, Germany. Hans D. Schotten\orcidlink{0000-0001-5005-3635} is also affiliated with the Intelligent Networks (IN) Research Department, German Research Center for Artificial Intelligence (DFKI), $\mathrm{67663}$ Kaiserslautern, Germany. The corresponding author to whom all correspondence should be addressed is Mohammad Asif Habibi\orcidlink{0000-0001-9874-0047} (\texttt{asif@eit.uni-kl.de}).}
}

\maketitle
\begin{abstract}
Efficient flexibility and enhanced system scalability require improved network performance, reduced energy consumption, lower infrastructure costs, and effective resource utilization. Achieving these objectives necessitates architectural optimization and reconstruction of existing cellular networks. Network slicing has emerged as a key enabler and architectural solution for communication systems in 2020 and beyond. Traditionally, network operators provide various services to different customers through a single network infrastructure. However, with the deployment of network slicing, operators can divide the entire network into multiple slices, each with its own configuration and specific quality of service (QoS) requirements. In a slice-based network, each slice is treated as an independent logical network. This approach enables more efficient infrastructure utilization and resource allocation in terms of both energy and cost compared to traditional telecommunications networks. In this paper, we provide a comprehensive discussion of the concept and system architecture of network slicing, with particular emphasis on its business aspects and profit modeling. Specifically, we examine two dimensions of profit modeling, namely \textit{Own-Slice Implementation} and \textit{Resource Leasing for Outsourced Slices}. Furthermore, we discuss open research directions and existing challenges to motivate further advancements and encourage practical solutions for this emerging technology.
\end{abstract}

\IEEEoverridecommandlockouts
\begin{keywords}
5G, Network Slicing, Network Slice, Network Architecture, Standardization, Wireless Communication
\vspace{-2.5mm}
\end{keywords}

\IEEEpeerreviewmaketitle

\section{Introduction}\label{Sec:Introduction}
\vspace{-1.5mm}
\IEEEPARstart{T}{he} fifth-generation (5G) communication system is expected to address the service, consumer, and business demands of the post-2020 era. In addition to supporting a massive number of user equipments, increased network traffic, and enhanced quality of service (QoS) requirements for telephony and data services, 5G also enables support for a wide range of vertical industries, including healthcare, manufacturing, automotive, logistics, energy, environmental monitoring, and construction. These industries require diverse use cases and varying QoS requirements; therefore, a \textit{one-size-fits-all} network architectural approach is no longer sufficient to fully leverage 5G and beyond communication systems and services. To efficiently accommodate vertical use cases alongside increasing user demands over the same network infrastructure, the 5G communication system requires architectural optimization and reconstruction compared with current deployments. Network slicing represents a promising architectural solution to meet these requirements, as it enables network operators to partition the communication network in a structured, elastic, scalable, and automated manner.

The deployment of network slicing enables the operation of multiple logical networks over a single physical infrastructure, thereby reducing overall costs, lowering energy consumption, and simplifying network functions (NFs) compared to operating a single network for multiple use cases or business scenarios. In a slice-based network, each slice is assigned its own specific characteristics and is treated as an independent logical network. As a result, infrastructure utilization and resource allocation can be achieved in a more energy- and cost-efficient manner compared to traditional network architectures.

The implementation of network slicing in 5G communication systems introduces numerous technical challenges that must be addressed. In addition, several business and economic issues (such as total cost and revenue) require significant optimization and redesign to align with the new network architecture. Meanwhile, the demand for broadband multimedia services has been increasing rapidly. If this trend continues, the revenue of mobile network operators may soon be surpassed by the capital expenditure (CAPEX) and operating expenditure (OPEX) required to maintain and operate the network infrastructure. Therefore, total cost, expected revenue, and resource allocation in the context of network slicing represent important research topics that warrant further investigation.

In this paper, we provide a comprehensive discussion of the concept and system architecture of network slicing, with particular emphasis on its business aspects and profit modeling. Specifically, we examine two dimensions of profit modeling, namely \textit{Own-Slice Implementation} and \textit{Resource Leasing for Outsourced Slices}. Furthermore, we discuss the associated challenges and open research directions in order to provide new insights and encourage realistic solutions to existing problems in the implementation of 5G slicing.

The rest of this paper is organized as follows. Section II reviews the state of the art related to network slicing. Section III discusses the concept and system architecture of network slicing for 5G. Section IV presents the business aspects and the two dimensions of profit modeling. Section V outlines future research directions. Finally, Section VI concludes the paper and summarizes the main findings.

\section{Related Work}\label{Sec:Related Work}
\vspace{-1mm}
The concept of network slicing has been extensively studied in the literature. For instance, the authors in \cite{1608.00572} present a detailed end-to-end framework for the implementation of network slicing in 5G. The paper discusses the deployment of both vertical and horizontal slicing across the air interface, the radio access network (RAN), and the core network (CN). Furthermore, it examines how computation and communication resources can be horizontally sliced to and from virtual computation platforms in order to improve scalability, enhance device capabilities, and increase the end-user experience.

Moreover, \cite{7891795}, \cite{7561023}, and \cite{7888080} focus on the deployment of network slicing within the RAN architecture. The authors in \cite{7891795} analyze network slicing in a multi-cell RAN to support radio resource partitioning among different slices. The paper further proposes four types of RAN slicing approaches and provides a detailed comparison among them. In contrast, the authors in \cite{7561023} examine how network slicing may affect various aspects of the design and functionality of the RAN architecture in 5G. The study also discusses the RAN requirements for the implementation of network slicing. Furthermore, \cite{7888080} provides a comprehensive discussion on the deployment of network slicing in heterogeneous cloud-RAN (C-RAN) environments to improve throughput through the sharing of computation, communication, and storage resources.

\begin{figure}
\centering
\includegraphics[width=.5\textwidth]{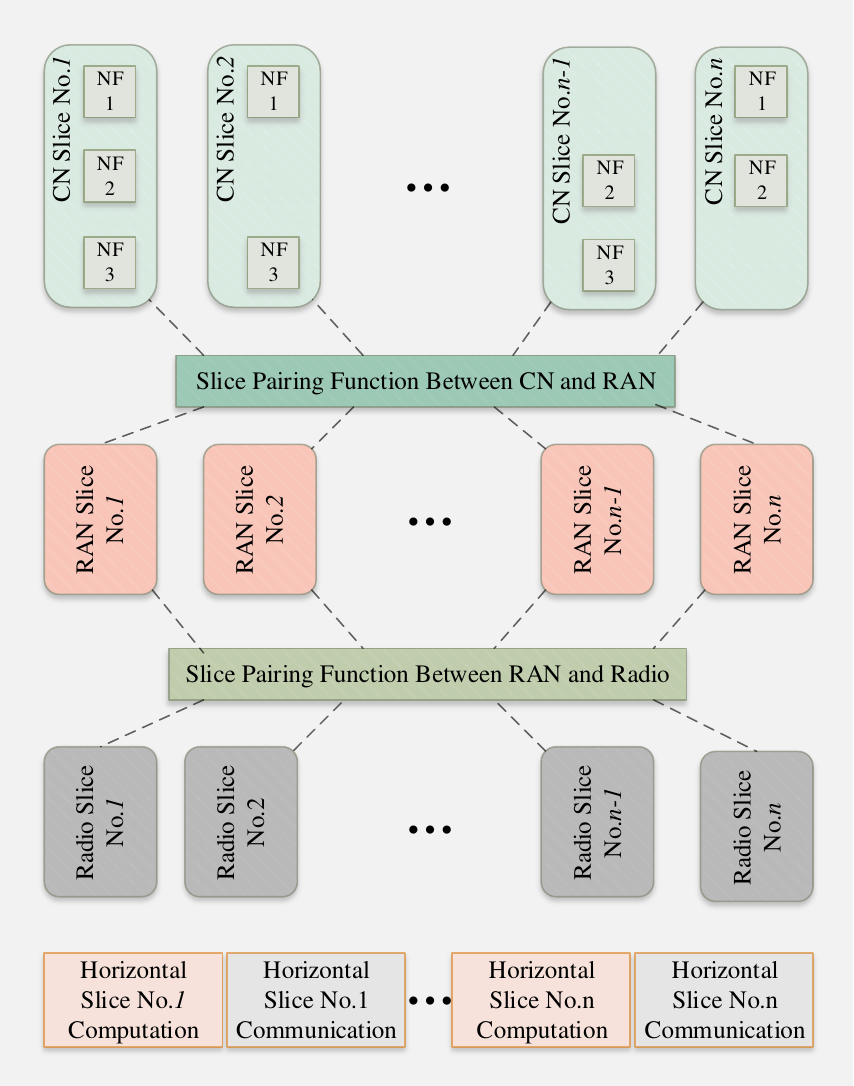}
\caption{The architectural framework of network slicing in 5G}
\label{fig:Network_Slicing}
\end{figure}

Furthermore, the authors in \cite{7926921} address several concepts related to network slicing, including resource allocation, virtualization technologies, orchestration processes, and isolation mechanisms. The paper provides a comprehensive discussion of software-defined networking (SDN) and network function virtualization (NFV), along with a deployment use case that considers network slicing through the integration of both NFV and SDN. The authors also highlight existing challenges and outline future research directions for the implementation of network slicing in 5G communication systems. Moreover, a comprehensive survey on the architecture of network slicing and its future research directions is presented in \cite{7926923}.

The Next Generation Mobile Networks (NGMN) Alliance introduced the network virtualization substrate (NVS), which enables an infrastructure provider to control the resource allocation of each virtual instance of an enhanced Node B (eNB) before customizing the scheduling policies for each virtual operator within the allocated resources. On the other hand, a heuristic-based admission control mechanism is proposed in \cite{7499297}, which dynamically allocates network resources to different slices in order to improve end-user satisfaction while considering the specific requirements of each slice.

One of the main objectives of network slicing is to optimize the profit modeling of traditional telecommunication networks. In order to increase overall revenue and reduce total network expenditure, a comprehensive study of the business and economic dimensions of network slicing is required. The authors in \cite{han2017modeling} analyze the profit generated by different slices operating over the same network infrastructure and further model network resource management. Meanwhile, \cite{bega2017optimising} addresses the design of an algorithm that allocates requests for network slices in a way that maximizes the total revenue of the network infrastructure provider.

However, in this paper, we provide a comprehensive discussion of the concept and system architecture of network slicing, with particular emphasis on its business aspects and profit modeling. Specifically, we examine two dimensions of profit modeling, namely \textit{Own-Slice Implementation} and \textit{Resource Leasing for Outsourced Slices}. Furthermore, we discuss open research directions to motivate new advances and propose realistic solutions for this emerging technology.

\section{Network Slicing: Concept and System Architecture}\label{Sec:Network Slicing: Concept and System Architecture}
\vspace{-1mm}
Network slicing, in its simplest form, refers to the use of virtualization and softwarization technologies, such as NFV and SDN, to design, partition, organize, and optimize the communication and computation resources of a physical network infrastructure into multiple logical networks in order to support a variety of services \cite{1608.00572}. With the deployment of network slicing, a single physical network infrastructure can be partitioned into multiple virtual networks, known as network slices. Each slice may have its own architecture, applications, and packet and signal processing capabilities, and is responsible for provisioning specific applications and services to particular end users and use cases. Examples of network slices include a slice dedicated to the remote control functions of a factory, a slice serving a utility company, and a slice designed to provide emergency healthcare services. A slice consists of both virtual network functions (VNFs) and physical network functions (PNF), which are appropriately composed to support and deliver services to end users.

The deployment of network slicing typically involves two main phases: creation and runtime \cite{7926921}. During the slice creation phase, an end user requests a slice from a network slice catalog, and the tenant provides the requested slice accordingly. In the runtime phase, the functional blocks within each slice—previously created during the creation phase—operate and deliver services according to the end user's requirements.

Each network resource, such as the NFV infrastructure and the functional blocks within a specific slice, should implement appropriate security mechanisms and ensure operation within expected parameters in order to prevent unauthorized access. This approach guarantees that faults or attacks occurring in one slice remain confined to that particular slice and do not propagate across slice boundaries. Furthermore, network slicing enables operators to provide new services and applications simply by deploying a new slice rather than rolling out an entirely new network, which helps reduce CAPEX and deployment time.

Network slices operate on a partially shared infrastructure. This infrastructure consists of both dedicated hardware—such as network elements in the RAN—and shared hardware resources, such as those provided by the Network Functions Virtualization Infrastructure (NFVI). NFs running on shared resources are typically instantiated in a customized manner for each network slice. However, this approach cannot always be applied to NFs that rely on dedicated hardware. Therefore, the design and identification of common functions represent important research directions in the context of network slicing in 5G and beyond.

There are two different concepts or scenarios for applying network slicing in 5G communication networks \cite{7529130}: slicing for the purpose of QoS provisioning and slicing for the purpose of infrastructure sharing. Both dimensions of network slicing are described as follows:

\begin{itemize}

\item \textbf{Slicing for QoS:} The basic idea is to create multiple network slices in order to offer different types of services to end users while ensuring specific QoS requirements within each slice. An example of this type of slicing is a slice created to provide services to a specific group of devices with particular QoS requirements, such as live video streaming or broadband connectivity for medical emergency response operations.

\item \textbf{Slicing for Infrastructure Sharing:} The fundamental idea of this network slicing scenario is to virtualize the RAN domain of a wireless network and share it among multiple operators. In this scenario, there is a slice owner and a slice tenant. The owner provides the slice to a tenant based on a service agreement. The tenant has overall control over both the functions and infrastructure of that slice. This concept of network slicing helps optimize the network cost model to increase overall revenue while simultaneously providing improved network scalability.

\end{itemize}

The purpose of network slicing in 5G mobile communication systems is to enable network operators to share infrastructure in a flexible and dynamic manner while efficiently managing network resources in the presence of an increasing number of devices and massive user traffic. A detailed discussion of the objectives and motivations behind the implementation of network slicing can be found in \cite{7529130}. Network slicing also helps mobile network operators simplify the creation, configuration, and operation of network services.

To enable efficient allocation of network resources, two-tier priorities are introduced in \cite{7499297}. The first tier is Inter-slice Priority, which refers to the prioritization among different network slices within a communication network. The priority of each slice is defined through an agreement between the slice owner and the tenant. The second tier is Intra-slice Priority, which refers to the prioritization among different users within a single network slice. These priorities are defined between the users and the service provider.

\begin{figure}
\centering
\includegraphics[width=.5\textwidth]{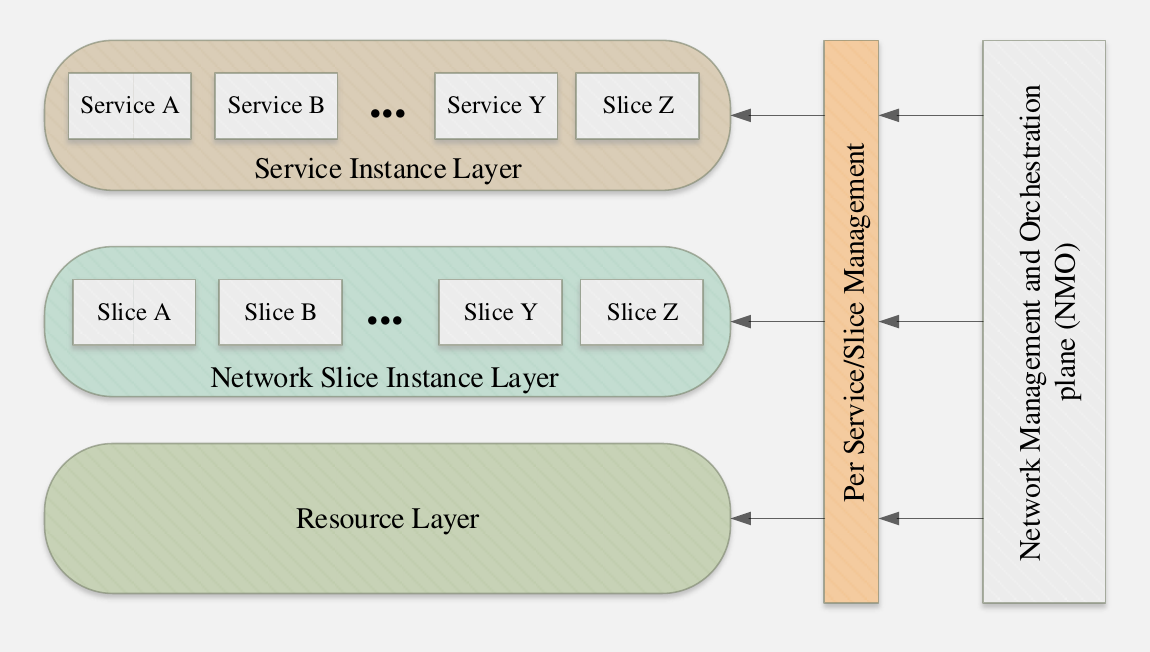}
\caption{Network slicing management and orchestration architecture}
\label{fig:Slicing_Management}
\end{figure}

Slicing in 5G communication networks can be deployed in two dimensions: Vertical Slicing and Horizontal Slicing. Vertical slicing primarily enables services for vertical industries and mainly focuses on the CN. In contrast, horizontal slicing aims to improve system performance and enhance the end-user experience, and mainly involves the RAN architecture. The two types of slicing are described in detail as follows \cite{1608.00572}:

\begin{itemize}
\item \textbf{Vertical Slicing:} The development and deployment of vertical slicing began in the late fourth generation (4G) and early 5G eras, and has primarily focused on the core domain of mobile networks. Mobile broadband networks are vertically sliced in order to support vertical industries and applications in a more cost-efficient manner. This approach segregates the traffic of vertical industries from the general broadband services of the mobile network, thereby simplifying traditional QoS challenges.

\item \textbf{Horizontal Slicing:} The increasing number of user equipments and the massive amount of traffic generated at the edge of mobile networks extend network slicing from the core domain to the RAN and air interface, a concept referred to as horizontal slicing. It is designed to accommodate emerging trends such as scaling system capacity, enabling cloud computing, and offloading computation from devices at the edge of mobile networks. Horizontal slicing enables resource sharing among network nodes and devices. For example, highly capable network nodes can share their resources—such as communication, computation, and storage—with less capable devices, thereby enhancing overall network performance.
\end{itemize}

Both vertical and horizontal slicing operate independently of each other. In vertical slicing, end-to-end traffic flows between the CN and user devices. In contrast, in horizontal slicing, traffic is typically exchanged locally between the two endpoints of a network slice, for example, between a portable device and a wearable device \cite{1608.00572}. In vertical slicing, each network node deploys similar functions across different network slices. However, in horizontal slicing, new functions may be introduced or instantiated at specific network nodes.

Fig.~\ref{fig:Network_Slicing} illustrates the concept and system architecture of network slicing. The architecture consists of CN slice subnets, RAN slice subnets, and radio slice subnets. Each CN slice is constructed from a set of NFs; some NFs may be shared across multiple slices, while others are tailored to specific slices. The architecture includes at least two slice pairing functions that connect these slices. The first pairing function links CN slices with RAN slices, while the second connects RAN slices with radio slices. The pairing function routes communication between a radio slice and its corresponding CN slice in order to support specific services and applications. The pairing function between RAN and CN slices may be static or semi-dynamic to achieve the required network functionality and communication. The mapping among radio, RAN, and CN slices can follow a 1:1:1 or 1:M:N relationship. This means that a radio slice may utilize multiple RAN slices, and a RAN slice may connect to multiple CN slices.

The end-to-end slicing architecture shown in Fig.~1 represents a logical decomposition of network slicing and takes specific network domain functions, such as the CN and RAN domains, into account. From an operational perspective, the NGMN Alliance defines the network slicing concept as consisting of three layers: the Service Instance Layer, the Network Slice Instance Layer, and the Resource Layer. Each of these layers is described below and illustrated in Fig.~\ref{fig:Slicing_Management} \cite{alliance2016description}.

\begin{itemize}
\item \textbf{Service Instance Layer:} This layer represents end-user and business services that are expected to be supported by the network. Each service is represented by a \textit{Service Instance}. These services may be provided either by the network operator or by a third party.

\item \textbf{Network Slice Instance Layer:} An operator uses a \textit{Network Slice Blueprint} to create a \textit{Network Slice Instance}. The slice instance provides the network characteristics required by a service instance. A network slice instance may be shared across multiple service instances provided by an operator. It may consist of none, one, or multiple \textit{Sub-network Instances}, which may also be shared with other slice instances. A \textit{Sub-network Blueprint} is used to create a sub-network instance that forms a set of NFs running on physical or logical resources.

\item \textbf{Resource Layer:} The actual PNFs and VNFs are used to implement a network slice instance. At this layer, network slice management functions are handled by the resource orchestrator, which is composed of an NFV Orchestrator (NFVO) and application resource configurators.
\end{itemize}

The Network Management and Orchestration (NMO) plane, shown on the right-hand side of Fig.~\ref{fig:Slicing_Management}, provides orchestration and management services for the three layers described above. The NMO functions must support orchestration and management services at a per-slice level.

\section{Network Slicing for Profit Modeling}
\vspace{-1mm}
Besides the improved scalability and flexibility provided by network slicing, it is also important to discuss the business dimension of applying network slicing from the operator's perspective. In traditional networks, the operator’s costs -- particularly in terms of CAPEX and OPEX -- are relatively high compared to the total generated revenue. There are several reasons for these higher costs, the most significant being the underutilization of network resources.

In traditional network architectures, the network operator provides a common pool of network resources for the general use of all applications. However, depending on their performance requirements, different applications may exhibit highly specific resource utilization characteristics. Consequently, under traditional architectures, it is often necessary to reserve excessive amounts of certain network resources for specific use cases, even when the actual demand is relatively low. As a result, many network resources remain underutilized, leading to a low overall network utilization rate.

In contrast, with the deployment of network slicing in 5G communications, network operators are able to efficiently analyze the operational costs and the revenue generated from each individual network slice within the network. Based on this revenue analysis, network operators can allocate specific bundles of network resources to different network slices, making resource management more structured, flexible, and efficient. As a result, the same network infrastructure can be utilized to seamlessly provide more and better services, thereby generating additional revenue without increasing CAPEX.

Consequently, network providers require new algorithms to adapt to the new architecture and maximize revenue. To achieve this goal, a comprehensive review of the telecommunications regulatory framework is also necessary. In addition, innovative pricing strategies, new mechanisms for cost sharing, and standardized solutions that support interoperability in multi-vendor and multi-technology environments must be explored. Furthermore, inter-operator network sharing and cooperative slicing concepts can be effectively implemented by optimizing the network cost model to increase overall revenue while simultaneously enhancing network scalability.

\begin{figure}
\centering
\includegraphics[width=.5\textwidth]{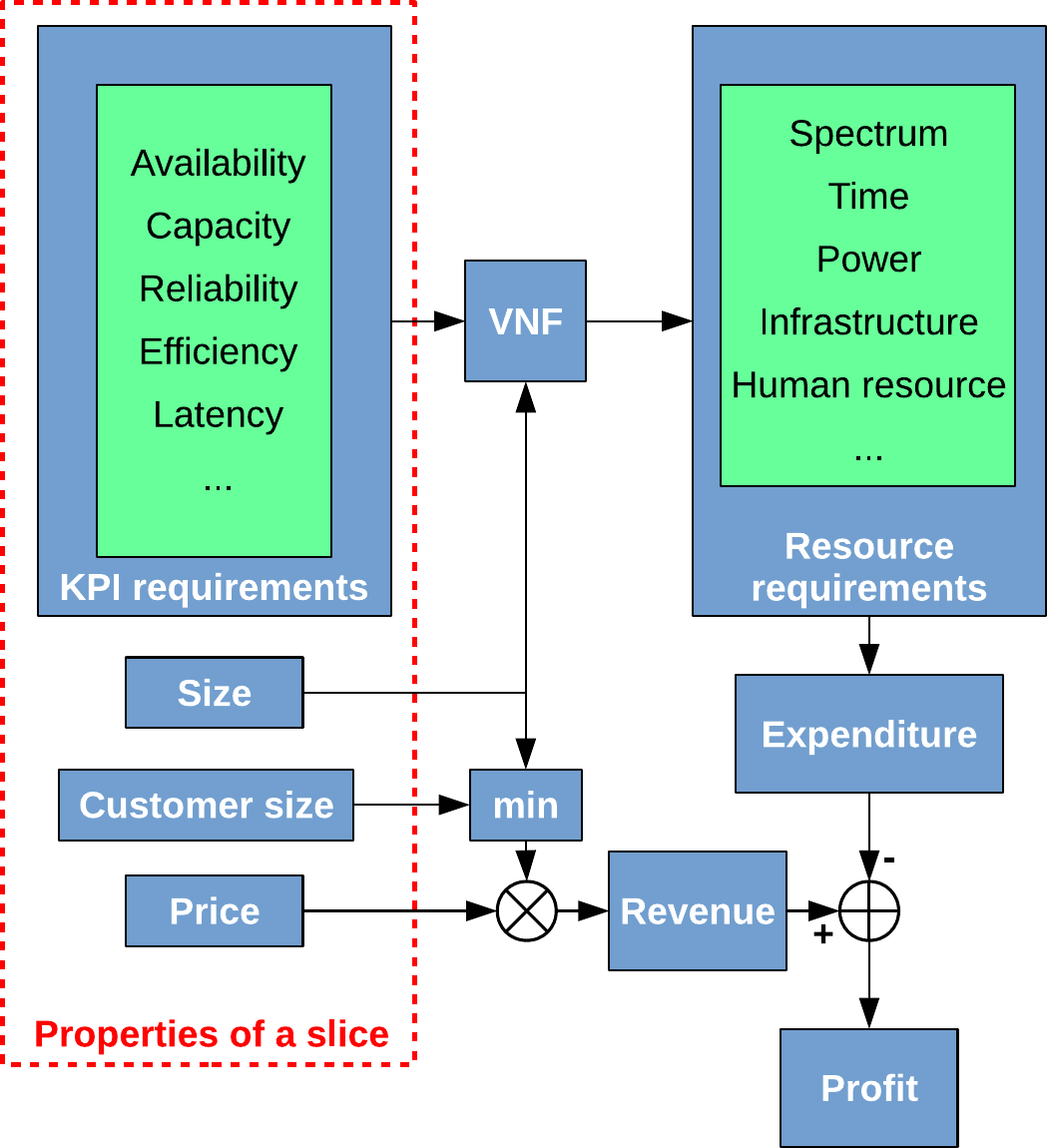}
\caption{Mapping chain that converts system properties into profit}
\label{fig:Profit_Slicing}
\end{figure}

Network cost models in traditional networks are typically based on CAPEX and OPEX, which are estimated according to factors such as the number of base stations (BSs), transmission power, and traffic volume. However, this methodology is no longer suitable for sliced networks. Since network resources can be shared among multiple network slices and the slicing scheme may vary across different resources, OPEX cannot be reliably estimated for the entire physical network as a whole. Therefore, a novel slice-oriented cost model is required. As discussed earlier, each network slice is designed to support a particular predefined service, which encompasses a group of use cases with similar QoS requirements. Consequently, a slice can be uniquely characterized by a specific set of key performance indicator (KPI) requirements.

Since 5G networks are expected to support network slices implemented by both traditional mobile network operators (MNOs), which own the network infrastructure and resources, and virtual MNOs (tenants), which rent infrastructure and network resources from traditional MNOs to implement and deliver services, two distinct business models emerge in 5G networks, leading to two corresponding profit models.

\begin{itemize}
\item \textbf{Own-Slice Implementation:} When an MNO implements a slice using its own network infrastructure and resources, it possesses prior knowledge about the slice, including the VNF scaling characteristics, implementation costs, customer demand for the slice service, and the service charge rate. Therefore, the operator can estimate the revenue and expenditure generated by the slice for an arbitrary slice size, as illustrated in Fig.~\ref{fig:Profit_Slicing}. By appropriately scaling the sizes of different slices, the MNO can optimize the overall network profit within the limits of its available resource pool. This process can be formulated as an optimization problem in resource allocation. This business model and profit framework have been discussed in \cite{han2017modeling}.

\item \textbf{Resource Leasing for Outsourced Slices:} Alternatively, an MNO can generate revenue by leasing its network resources in predefined bundles to tenants for their network slice implementation. Different resource bundles may be offered according to slice characteristics such as service type and performance elasticity. In this scenario, lease requests are submitted by tenants, while the MNO decides whether to accept or reject each request. Once a lease is confirmed, the MNO cannot flexibly scale or terminate the tenant slice. Nevertheless, by establishing an appropriate decision mechanism based on the statistical characteristics of incoming requests, the MNO can still optimize the long-term overall revenue within the limits of its available resource pool. An example of this business model and the corresponding optimization problem is presented in a detailed manner in \cite{bega2017optimising}.
\end{itemize}

A detailed discussion and comparison of these two dimensions of cost modeling are beyond the scope of this paper. However, this topic can be further explored in future work along with the research directions outlined below.

\section{Open Research Challenges and Future Directions}
\vspace{-1mm}
Network slicing is still at an early stage of development; therefore, further enhancements and studies are required for it to become a mature technology and to be widely adopted across various domains of emerging 5G systems. Despite the significant advantages that network slicing brings to 5G systems, several challenges remain. In this section, we identify key research problems and future directions in the field of network slicing that need to be investigated for its full implementation across all components of 5G \cite{1608.00572,7926921,7926923}.

\textbf{1 — Business and Profit:} The integration of multiple slices, each providing services at different stages and targeting specific end users, along with partnerships between several network operators through infrastructure sharing, creates new challenges related to total network investment, service level agreements (SLAs) between slice owners and tenants, and expected revenue generation. Considering such an integrated, business-oriented approach, new economic strategies and profit models must be thoroughly analyzed and developed to meet the requirements of 5G networks. To achieve this goal, an in-depth study of the existing telecommunications regulatory framework is required. In addition, innovative pricing strategies, cost models for infrastructure sharing, SLAs between slice owners and tenants, and revenue generation mechanisms should be carefully investigated and further standardized.

\textbf{2 — Security:} Existing and proposed open interfaces in network slicing that support network programmability introduce new potential attack vectors in software-defined networks. These concerns represent major barriers to the deployment of 5G networks and call for extensive research and the development of a multi-level security framework consisting of both policies and mechanisms. Such a framework should include dynamic threat detection, user authentication, accounting management, and remote attestation.

\textbf{3 — Management:} Despite the improved dynamicity and scalability that network slicing brings to 5G communication systems, network management and orchestration in multi-tenant environments remain major challenges. In order to dynamically assign network resources to different slices, the optimization policies governing the resource orchestrator must be able to handle situations where resource demands vary. To achieve this goal: i) effective cooperation between slice-specific management functional blocks and the resource orchestrator is required; ii) all management policies must be automatically validated; and iii) efficient resource allocation algorithms and conflict resolution mechanisms must be designed and implemented at each abstraction layer.

\textbf{4 — Performance:} The 5G communication system consists of multiple virtual networks, different radio access technologies, and diverse QoS requirements operating over the same network infrastructure. When network slices are deployed, network performance analysis and QoS measurement become more challenging tasks. Therefore, comprehensive studies are required to develop effective solutions for dynamic performance monitoring and network analysis while considering both time and cost constraints.

\textbf{5 — Standardization:} The standardization process for network slicing is still in its initial phase and has primarily focused on vertical slicing. A wide range of studies on network slicing are currently being conducted by various research initiatives and projects, including NGMN Alliance, 5G Novel Radio Multiservice adaptive network Architecture (NORMA), the Co-Funded Framework, Wireless World Research Forum (WWRF), Third Generation Partnership Project (3GPP), and 5G Infrastructure Public Private Partnership (5G PPP). Across these initiatives, network slicing is widely considered one of the fundamental requirements for 5G communication systems. The current stage of network slicing development mainly focuses on defining its concept, system architecture, requirements across different network segments, and the impact of slicing on overall network architecture. Considering these diverse research directions, comprehensive global standardization is required for the effective deployment of network slicing in emerging 5G communication systems. The full standardization process of network slicing is expected to be finalized in the context of 3GPP Release 15 and beyond.

\textbf{6 — RAN Virtualization:} Core network slicing has already been extensively investigated, and numerous studies have focused on the efficient slicing of this particular domain. However, one of the major challenges for further network virtualization lies in the RAN of 5G systems. Since 5G networks incorporate multiple access technologies, it is essential for RAN virtualization solutions to accommodate these diverse technologies. This situation introduces an additional challenge for the RAN, as it remains unclear whether multiple access technologies can be multiplexed over the same hardware platform or whether each technology will require dedicated hardware. Addressing these questions requires further investigation through targeted studies and research in this area.

\section{Conclusions}
\vspace{-1mm}
In this paper, we provided a comprehensive discussion of the concept and system architecture of network slicing in 5G systems, with particular emphasis on its business aspects and profit modeling. We examined two dimensions of profit modeling, namely Own-Slice Implementation and Resource Leasing for Outsourced Slices. Furthermore, we addressed existing challenges and identified open research directions in the field of network slicing in order to provide new insights and encourage the development of realistic solutions.

In future work, we intend to extend several aspects of this study. From the system architecture perspective, we plan to investigate the development and implementation of horizontal slicing over RAN architectures and end-user devices. From the profit modeling perspective, we aim to design and develop more realistic models for resource costs and service revenue with detailed parameters. In addition to these directions, several other important research challenges have been discussed in the previous section of this paper.


\bibliography{ref/mypaper01.bib} 
\bibliographystyle{ieeetr}
\vspace{-1.5mm}


\section*{Author Biographies}
\vskip 0pt plus -1fil

\begin{IEEEbiography}[{\includegraphics[width=1in,height=1.25in,clip,keepaspectratio]{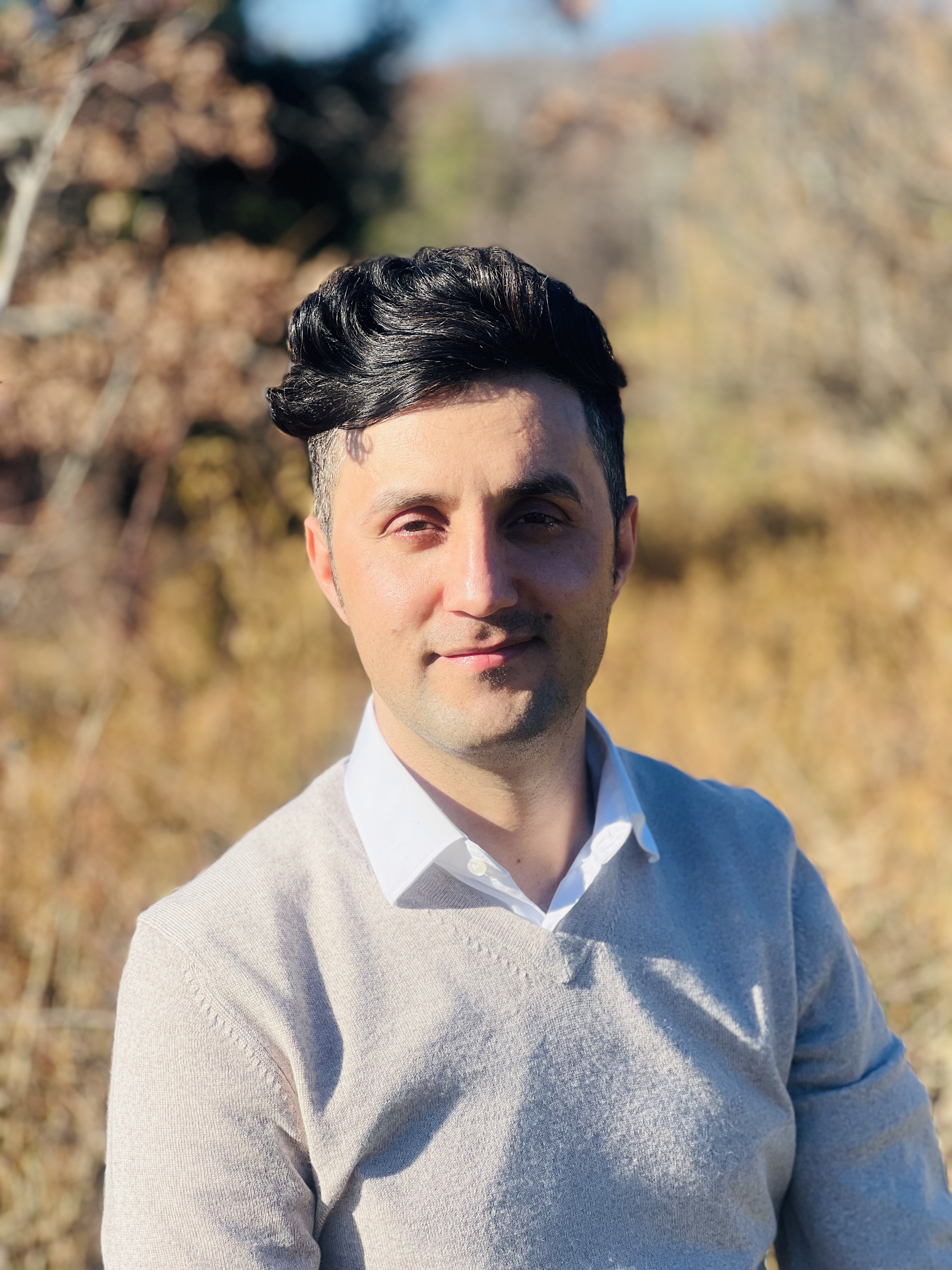}}]{Mohammad Asif Habibi} received his B.Sc. degree in telecommunication engineering from Kabul University, Afghanistan, in 2011. He obtained his M.Sc. degree in systems engineering and informatics from the Czech University of Life Sciences, Czech Republic, in 2016. Since January 2017, he has been working as a research fellow and Ph.D. candidate at the Division of Wireless Communications and Radio Navigation, Technische Universit\"at Kaiserslautern, Germany.  From 2011 to 2014, he worked as a radio access network engineer for HUAWEI. His main research interests include network slicing, network function virtualization, resource allocation, machine learning, and radio access network architecture.
\end{IEEEbiography} 
\vskip 0pt plus -1fil

\begin{IEEEbiography}[{\includegraphics[width=1in,height=1.25in,clip,keepaspectratio]{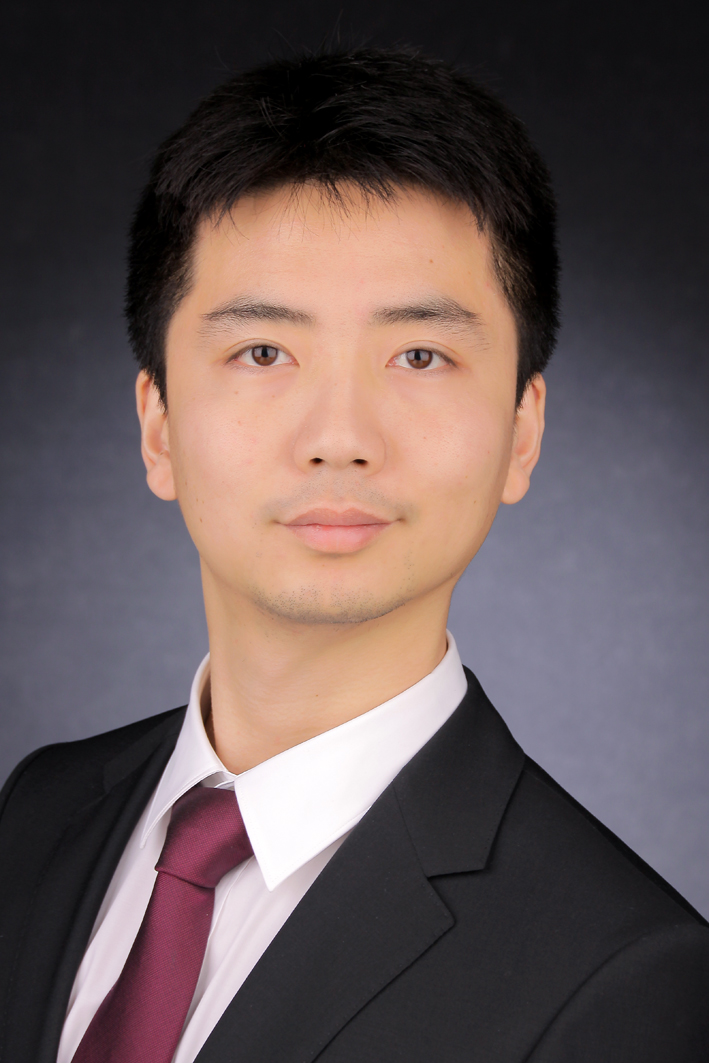}}]{Bin Han} (M'15 -- SM'21 ) received in 2009 his B.E. degree from Shanghai Jiao Tong University, in 2012 the M.Sc. from Technische Universit\"at Darmstadt, and in 2016 the Ph.D. (Dr.-Ing.) degree from Karlsruher Institute f\"ur Technologie. He joined Rheinland-Pf\"alzische Technische Universit\"at (previously known as Technische Universit\"at Kaiserslautern) in 2016, and is now a Senior Lecturer at its Division of Wireless Communications and Radio Navigation. Researching in the area of wireless communication and networking, he has authored over around 50 research papers and book chapters, and participated in multiple EU collaborative research projects.
\end{IEEEbiography} 

\begin{IEEEbiography}[{\includegraphics[width=1in,height=1.25in,clip,keepaspectratio]{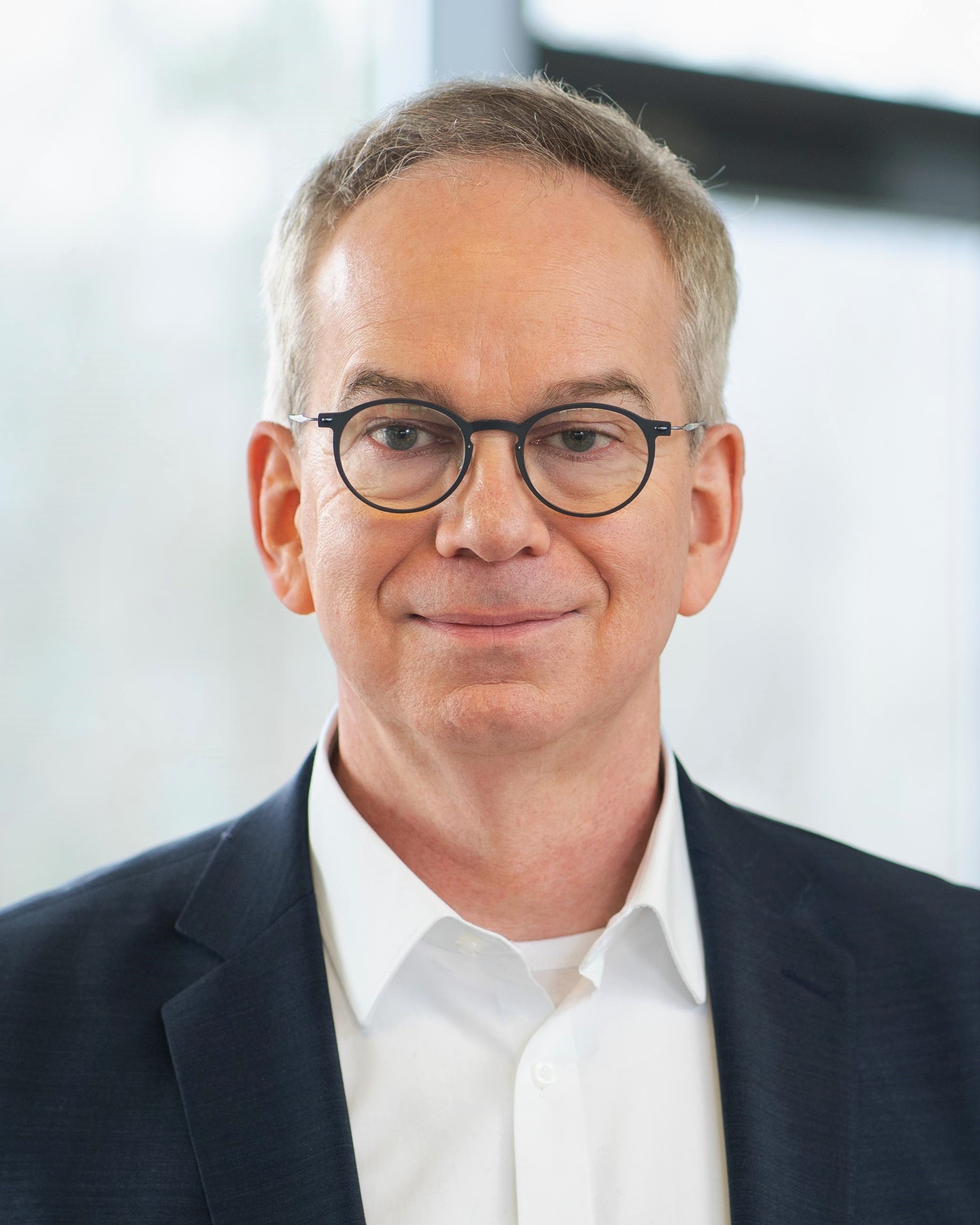}}]{Hans D. Schotten} received the Diploma and Ph.D. degrees in electrical engineering from the Aachen University of Technology, Germany, in 1990 and 1997, respectively. Since August 2007, he has been a full professor and head of the Division of Wireless Communications and Radio Navigation at Technische Universit\"at Kaiserslautern. Since 2012, he has also been Scientific Director at the German Research Center for Artificial Intelligence, heading the Intelligent Networks department. He was a senior researcher, the project manager, and the head of the research groups at Aachen University of Technology, Ericsson Corporate Research, and Qualcomm Corporate R\&D. During his time at Qualcomm, he has also been the Director for Technical Standards and Coordinator of Qualcomm’s activities in European research programs.
\end{IEEEbiography}

\end{document}